\documentclass[10pt,a4paper]{article}
\usepackage[a4paper, total={6in, 8in}]{geometry} 

\usepackage[latin1]{inputenc}
\usepackage{amsmath}
\usepackage{amsfonts}
\usepackage{amssymb}
\usepackage{graphicx}

\usepackage{amsmath}
\usepackage{graphicx,psfrag,epsf}
\usepackage{enumerate}
\usepackage{natbib}
\usepackage{url} 
\usepackage{bbm}
\usepackage{subcaption} 
\usepackage[linesnumbered,ruled]{algorithm2e}
\usepackage{rotating}
\usepackage{booktabs}
\usepackage{xcolor}
\usepackage{hyperref}
\usepackage{float}
\usepackage{verbatim}
\usepackage[normalem]{ulem}
\useunder{\uline}{\ul}{}
\usepackage{dirtytalk}

\newcommand{\NPF}{$\mbox{NP-FOCuS}$} 
\newcommand{\arr}{\xleftarrow{}}

\newtheorem{theorem}{Theorem}
\newtheorem{lemma}[theorem]{Lemma} 
\newtheorem{proposition}[theorem]{Proposition}

\author{Gaetano Romano, Idris Eckley, Paul Fearnhead}
\title{A Log-Linear Non-Parametric Online Changepoint Detection Algorithm based on Functional Pruning}

\begin{document}

\maketitle

\begin{abstract}
Online changepoint detection aims to detect anomalies and changes in real-time in high-frequency data streams, sometimes with limited available computational resources. This is an important task that is rooted in many real-world applications, including and not limited to cybersecurity, medicine and astrophysics.
While fast and efficient online algorithms have been recently introduced, these rely on parametric assumptions which are often violated in practical applications.
Motivated by data streams from the telecommunications sector, we build a flexible nonparametric approach to detect a change in the distribution of a sequence. Our procedure, NP-FOCuS, builds a sequential likelihood ratio test for a change in a set of points of the empirical cumulative density function of our data. This is achieved by keeping track of the number of observations above or below those points. Thanks to functional pruning ideas, NP-FOCuS has a computational cost that is log-linear in the number of observations and is suitable for high-frequency data streams. In terms of detection power, NP-FOCuS is seen to outperform current nonparametric online changepoint techniques in a variety of settings. We demonstrate the utility of the procedure on both simulated and real data.
\end{abstract}

\section{Introduction}

One of the contemporary challenges in time-series analysis is to detect changes in some measurable properties of a process. This task finds its roots in a plethora of applications spanning numerous fields including engineering \citep{alvarez2020flight, henry2010fault}, neuroscience \citep{jewell2020fast}, genomics \cite{nicolas2009transcriptional} and astrophysics \citep{fridman2010method,fuschino2019hermes}. In the previous decade, we saw many offline changepoint procedures appearing in the statistical literature. A common approach for detecting a single change is to scan over all possible locations for a change, and apply a generalised likelihood ratio (GLR) test for a change, with evidence for a change being the maximum of the GLR test statistics \cite[]{fearnhead2022detecting}. The GLR procedure has been proven to be asymptotically optimal \citep[][]{basseville1993detection}, as it searches over all possible parameters of the unknown pre and post-change distributions. This approach includes, as a special case, the popular CUSUM method for detecting a change in mean, and can be extended to detect multiple changepoints by using binary segmentation methods \cite[]{scott1974cluster,fryzlewicz2014multiple,kovacs2020seeded} or by maximising a penalised likelihood \cite[]{fearnhead2020relating}; see \cite{cho2021data} for a recent review of this area.

In contrast to the offline setting, where the data is first collected and later analyzed \emph{a posteriori}, one of the modern challenges is to detect a change within a data stream in real time. We find many applications in need of an online procedure. These include  control of industrial processes \citep{pouliezos2013real}, or monitoring of computer networks \citep{tartakovsky2005nonparametric}, social networks \citep{chen2019sequential} and telecommunication devices \citep{austin2023online}. 

Many challenges arise in analysing data online. An online procedure should be able to process observations in real-time, as quickly as they arrive, in order to avoid memory overflow and result in a delayed evaluation. This can be particularly valuable in settings where limited computational resources are available, or in the high-frequency domain.

As noted by \citep{ross2015parametric}, one way of performing an online analysis is to collect data in batches and analyse those through an offline algorithm. However, such an approach can be sensitive to the batch size. If this is too small, then we may be unable to detect small changes, while if it is large it can lead to delayed detection of bigger changes.
Alternatively, observations can be processed on the go in a sequential fashion. At each new observation, a decision is made on whether a change has occurred, or not, based on the new data point and on past information. 


The sequential CUSUM approach, and more generally, sequential  GLR approaches, demonstrate excellent statistical properties   \cite[]{wang2022sequential,yu2020note}. However, they can be computationally inefficient. A naive computation of the GLR test, in fact, involves, at time $n$, considering $\mathcal{O}(n)$ possible locations for a change. I.e., the algorithm has a computational cost per iteration that increases linearly, which is impracticable for online applications. 

Recently, \cite{romano2021fast} presented the FOCuS procedure, a fast algorithm to perform the sequential CUSUM test, that decreases the computational complexity from $\mathcal{O}(n)$ to $\mathcal{O}(\log n)$ per iteration. 
This is to our knowledge the fastest way to solve the sequential CUSUM test exactly.
For example, the expected cost per iteration of FOCuS at a given time, say at iteration one million, is roughly equal to the cost of iterating approximately 20 objects stored in memory to find the global maximum of our statistics -- and thus it is suitable for high-frequency online applications.

As previously mentioned, the CUSUM and GLR tests in general rely on parametric assumptions that are often hard to meet in real-world applications. 
For instance, with an industrial collaborator, we found that certain data streams arising in the telecommunications sector do not follow common distributions, nor do they satisfy the usual parametric assumptions. In monitoring operational metrics from network devices, for example, we often deal with nonstandard distributions with multiple modes, outliers or heavy tails. In many cases, the nature of the change is unknown \emph{a priori}, see for example the sequences from Figure \ref{fig:application_example}. In Section \ref{sec:application} we present one example of a contemporary telecommunication application, where the aim is to monitor the operational performance of devices on an optical cable network. In those scenarios, a test for Gaussian change-in-mean would, in fact, be prone to either a missed detection, in the case of a change of a different nature, or to false positives, in case of misspecification of the underlying noise process.

\begin{figure*}
    \centering
    \includegraphics[width=.7\linewidth]{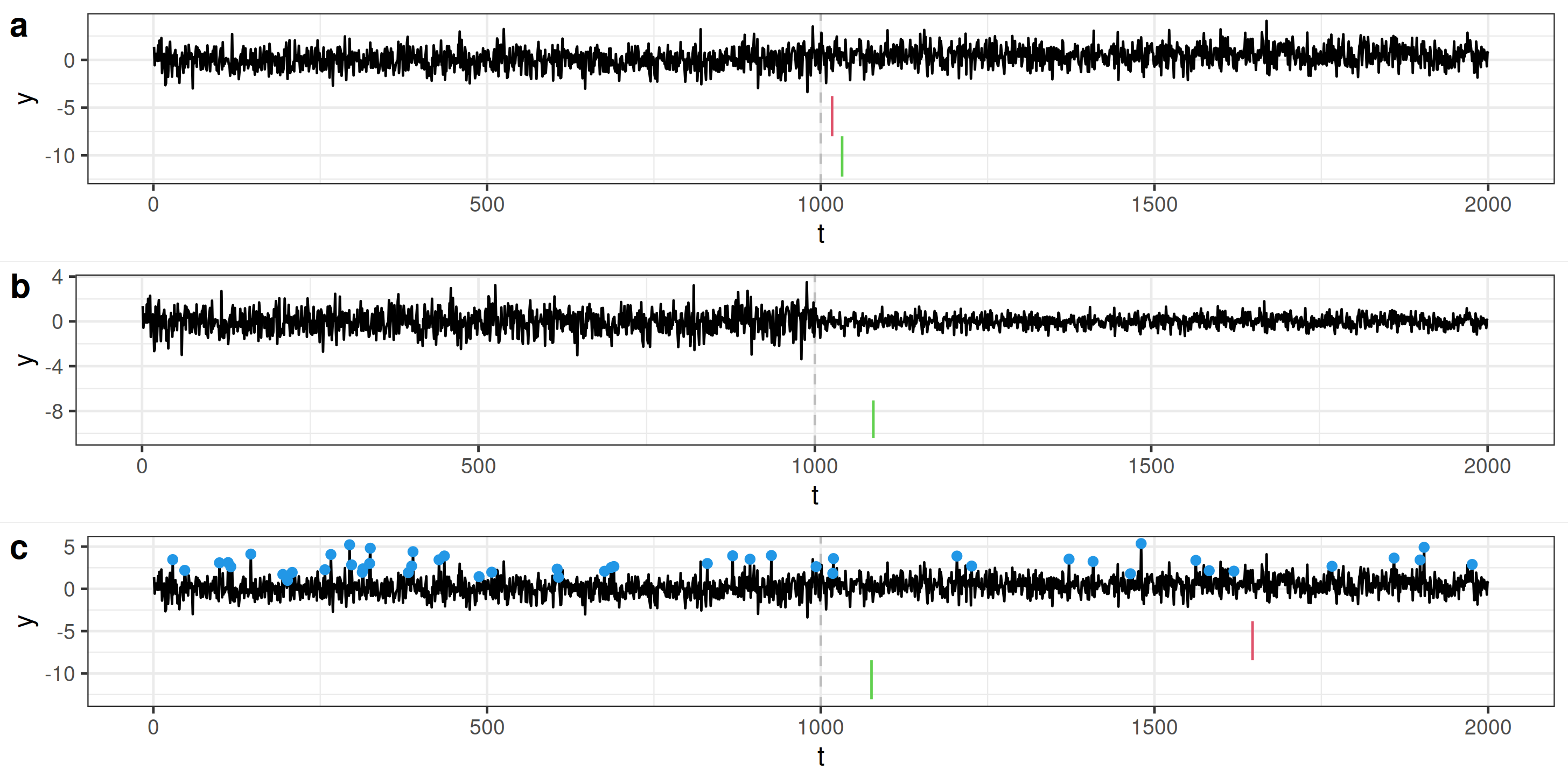}
    \caption{Detection delays of Gaussian FOCuS (in red) and NP-FOCuS (in green) for 3 sequences of length 2000 with a change at $t = 1000$.  In a, a Gaussian change-in-mean example, with a jump size of 0.25. In b, a change-in-variance from variance $\sigma^2 = 1$ to $\sigma^2 = 0.25$. Lastly, in c, we find the same scenario in a with some of the observations shifted by a value of 3 at uniformly random sampled locations (blue dots).}
    \label{fig:intro_example}
\end{figure*}

To better outline the limitations of a parametric approach in online changepoint detection, we show a simple introductory example. Let us compare the Gaussian FOCuS from \cite{romano2021fast} with its non-parametric counterpart, NP-FOCuS, the methodology that will be introduced in this paper. In Figure \ref{fig:intro_example} we study both the statistics on 3 different simulated changepoint scenarios. Thresholds were tuned to achieve comparable average run lengths of 2000 observations under the null (the expected number of observations until we detect a change if a change is not present). This way, after placing a true change at 1000, we can have a fair comparison focusing on the detection alone. In Figure \ref{fig:intro_example}a, the simple Gaussian change-in-mean case, we notice how the Gaussian procedure achieves the fastest detection. This is because the procedure's assumptions hold. However in Figure \ref{fig:intro_example}b, the second example, the sequence now shows a change in variance. In this setting, the Gaussian change-in-mean has no power to detect the change, which is missed.
Lastly, in Figure \ref{fig:intro_example}c, we find the same series as in the first example, but some observations are shifted up by 3. Under this scenario, to achieve the same run length of 2000 observations under the null, the threshold for the Gaussian FOCuS procedure needs to be inflated. This results in a slower detection than its non-parametric counterpart. 

Whilst there are numerous offline non-parametric approaches in the current literature \citep[including and not limited to][]{pettitt1979non, zou2014nonparametric, Haynes, matteson2014nonparametric,chen2023graph}, it is only more recently that there has been substantial interest on online approaches. This includes \cite{Ross}, who propose a control-chart approach that allows for the detection of changes in scale or location, or both, when the underlying distribution of the process is unknown, and \cite{shin2022detectors} who propose a general framework for detecting changes in measurable properties of a process.
More generally there are methods that choose a number of univariate summaries of data and monitor for a change in any of these summaries \cite[]{kurt2020real,kerivenNEWMA}, kernel-based methods \cite[]{flynn2019change,huang2014high} or related methods that use information on distances between data points \cite[]{chen2019sequential, chu2022sequential, chen2020high}. These approaches  can be applied in multivariate settings, but their statistical efficiency depends on choosing appropriate choice of summary, kernel or distance that had power to detect the type of change that occurs. These methods are often not invariant to transformations of the data. By comparison, methods based on ranking information \cite[e.g.][]{Hawkins}, are simple to apply for univariate data and are invariant to any monotone transformation of the data. This is also the case for methods that look for a change in the distribution function for univariate data, which is the approach we take in this paper.

One example of such a method is the NUNC algorithm \cite[]{austin2023online}, which tests for a change in the empirical cumulative distribution function (eCDF) within a rolling window.
As with any other moving window method, however, NUNC's performance is extremely dependent on the size of the window. 
For example, a window too small would end up missing changes of a smaller size, whilst a window too big might result in longer detection delays and an increased computational cost.

Our procedure, NP-FOCuS, tries to mitigate such issues by building a GLR test \citep[extending ideas from][]{romano2021fast} for the non-parametric procedure NUNC of \cite{austin2023online}. Our procedure maps the change-in-distribution problem into a Bernoulli GLR test. This is achieved by evaluating the eCDF at a fixed point and keeping track of the number of observations above or below that point. 
We then extend the FOCuS algorithm so that it can apply to the Bernoulli GLR test, and show theoretically that it has the same strong computational properties as the original FOCuS algorithm. We perform this test for a grid of quantiles, and then merge the test statistic values. In practice our final procedure monitors both the sum of the test statistics across the quantiles, which can detect small shifts in the distribution, and also the maximum of the test statistics, which can detect a larger change in just one part of the distribution.
Our approach only assumes that the observations are \textit{i.i.d.}. However, we also demonstrate empirically that even if such a hypothesis is violated we still retain the strong performance over competitor methods.

The paper is structured as follows. In Section \ref{sec:procedure} we introduce the NP-FOCuS procedure, starting from related work, to the introduction of a novel functional pruning recursion. In Section \ref{sec:complexity} we provide some guarantees on the computational complexity of NP-FOCuS. In Section \ref{sec:simulations} we evaluate empirically performances of NP-FOCuS, with a comparison to other methods in a simulation study. In Section \ref{sec:application} we evaluate the method on some real-word data streams.

Open-source code, making the methods developed in this paper accessible, will be made available via an appropriate repository in due course.

\section{A Functional Representation for the Nonparametric Cost Function}\label{sec:procedure}

\subsection{Overview of Problem and Approach}

We aim to build an online non-parametric changepoint detector by monitoring the empirical Cumulative Density Function (eCDF) of a stream of observations. This approach has been shown to work well in an offline setting by \cite{zou2014nonparametric, Haynes}.

Let $p \in \mathbb{R}$ and let $F_{1:n}(p)$ be the unknown CDF for a series of independent real-valued observations $y_1, \dots, y_n$, and let $\hat{F}_{1:n}(p)$ be its eCDF: 
\[
    \hat{F}_{1:n}(p) = \frac{1}{n} \sum_{t = 1}^n \left[ \mathbb{I}(y_t \leq p) 
    \right],
\]
where $\mathbb{I}$ is an indicator function. For an independent stream of observations, for a fixed $p$, the eCDF follows a binomial distribution:
\[
    n\hat{F}_{1:n}(p) \sim \mathrm{Binom}(n, \theta),
\]
with $\theta = F_{1:n}(p)$. 
A simple approach to detect a change in $\hat{F}_{1:n}(p)$ would be to record which data points are above $p$ and build a likelihood-ratio test for the Bernoulli data.
That is, for a fixed $p$, we know that the likelihood for a segment is given by
\begin{equation}\label{eq:log-lik}
\begin{split}
    &\mathcal{L}(y_{\tau_1+1:\tau_2};\ \theta, p) = (\tau_2 - \tau_1) \times \\ 
    &\times \left[\hat{F}_{\tau_1+1:\tau_2}(p) \log \theta + (1-\hat{F}_{\tau_1+1:\tau_2}(p))\log(1-\theta) \right].
\end{split}
\end{equation}
The GLR test for a change in parameter from $\theta_0$ to $\theta_1$ is
\begin{equation}
    \max_{0 \leq \tau < n} \left[ - \mathcal{L}(y_{\tau+1:n}; \theta_0, p) +\mathcal{L}(y_{\tau+1:n}; \theta_1, p) \right].\label{eq:np-focus-LR}
\end{equation}
By writing $x_t = \mathbb{I}(y_t \leq p)$, this can be rewritten as 
\begin{eqnarray*}
     \lefteqn{\mathcal{L}(y_{\tau+1:n}; \theta_1, p) -\mathcal{L}(y_{\tau+1:n}; \theta_0, p)}  \\ & =&  \sum_{t=\tau+1}^n \left[ x_t \log\theta_1 + (1-x_t)\log(1-\theta_1) \right]\\
     & -& \sum_{t=\tau+1}^{n} [x_t \log\theta_0 + (1-x_t)\log(1-\theta_0) ] 
\end{eqnarray*}
If both $\theta_0$ and $\theta_1$ are known, then the sequential procedure of \cite{Page} can be used to calculate the GLR test (\ref{eq:np-focus-LR}). 
In Section \ref{sec:change-in-rate} we show how to extend the FOCuS algorithm of \cite{romano2021fast} to calculate the GLR test if either only $\theta_0$ is known, or if neither $\theta_0$ nor $\theta_1$ is known.  We will call this procedure Ber-FOCuS. 

In practice, we would want to check changes in the entire eCDF rather than just at a single quantile value $p$. To achieve this we follow \citet{zou2014nonparametric} and \citet{Haynes} who suggest summing up the test statistic across a grid of values  $p_1, \dots, p_M$. 
Specifically, we approximate our eCDF on a fixed grid of $M$ quantile values $p_1, \dots, p_M$ computed over a probation period. To obtain NP-FOCuS we run the Ber-FOCuS routine independently for each quantile of our eCDF. To construct a global statistic we aggregate the quantile statistics through the sum and the maximum. In Section \ref{sec:aggregators} we will formally describe the full procedure.

\subsection{Detecting change-in-rate in a Bernoulli process}\label{sec:change-in-rate}

We focus on detecting a change in the rate parameter $\theta$ in a univariate stream of a Bernoulli process. Let $x_n \in \{0, 1\} \sim \text{Ber}(\theta)$ be a realization of a Bernoulli random variable with parameter $\theta$, with $\theta$ subject to a change. Assume the pre-change rate parameter, $\theta_0$, is known. For every observation we can obtain evidence for a post-change rate parameter $\theta_1$ via the likelihood ratio test: 
\begin{align}
    g(x_n, \theta_1) &= -2 \log\left( \frac{\theta_0^{x_n} (1-\theta_0)^{1-x_n}}{\theta_1^{x_n} (1-\theta_1)^{1-x_n}} \right) =\\
    &=2 \left [ x_n \log \left( \frac{\theta_1}{\theta_0} \right) + (1-x_n) \log\left(\frac{1-\theta_1}{1-\theta_0}\right)\right ]\label{eq:cost_ber}.
\end{align}
At time $n$ we have observed $x_1, \dots, x_n$, we can then employ \eqref{eq:cost_ber} and derive the test statistics: 
\[
\mathcal{Q}_{n, \theta_1} = \max_{0 \leq \tau < n} \ \sum_{t=\tau+1}^n g(x_t, \theta_1),
\]
where we maximise for all possible time-points of the change, $\tau$. Such a statistic is known as the sequential Page-CUSUM statistic \citep{Page}, and can be solved efficiently in constant time per iteration through the recursion:
\begin{equation}
\label{eq:page-recursion}
\mathcal{Q}_{n, \theta_1} = \max \left\{ 0, \ \mathcal{Q}_{n - 1, \theta_1} + g(x_n, \theta_1) \right\}.     
\end{equation}

This statistic is straightforward to compute for a known value of $\theta_0$ and a fixed value of $\theta_1$. As we do not know $\theta_1$, one would compute the statistics over a fixed grid of values for $\theta_1$ and choose the maximum value of those to maximise the power of a detection. However, this would introduce a trade-off between power and computational complexity.  Alternatively, \cite{xieWindowLimited} recently suggested  a way of approximating the Page-CUSUM cost by using a plug-in value for $\theta_1 = \hat{\theta}$ into the equation above, that is calculated recursively based on only recent data point. 

\begin{figure}
    \centering
    \includegraphics[width=.3\linewidth]{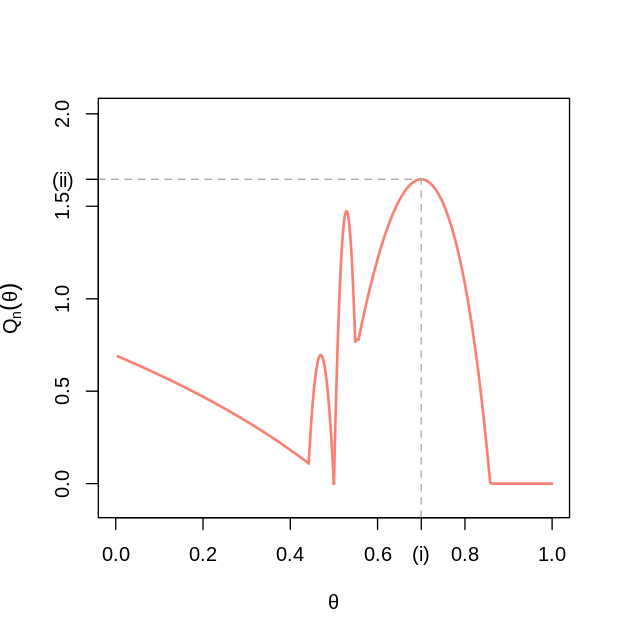}
    \caption{The functional cost $Q_n(\theta)$ at iteration $n$. Dotted lines illustrate (i) $\text{arg}\max_\theta Q_n(\theta)$ and (ii) the $\max_\theta Q_n(\theta)$, the highest possible value of our test statistics.}
    \label{fig:functional_cost_ill}
\end{figure}

Our approach is to calculate the GLR test statistic exactly for both the case where $\theta_1$ is unknown. The idea, based on \cite{romano2021fast}, is to solve \eqref{eq:page-recursion} simultaneously for all values of $\theta$ through the functional representation of the sequential test statistics $Q_n(\theta)$ for a post-change rate parameter $\theta$. That is, we have $Q_0(\theta)=0$ and: 
\begin{equation}
        \label{eq:functional-recursion}
        Q_n(\theta) = \max \left\{0, \ Q_{n - 1} (\theta) + g(x_n, \theta) \right\}
\end{equation}
for $n \geq 1$. 
The value of our test statistic will be  $\mathcal{Q}_n = \max_\theta Q_n(\theta)$. 
In this optimization lies the major computational contribution of NP-FOCuS. At each iteration of the algorithm we explicitly compute and store the full functional representation of our cost $Q_n(\theta)$, which will be a piecewise function (see Figure \ref{fig:functional_cost_ill}).
By maximizing over all possible values of $\theta \in \ (0, 1)$, which is simple as we can maximise each piecewise part of the function and take the maximum of these maxima, we can find the highest possible value for the Page-CUSUM statistics, and avoid altogether any choice of window $w$ or specify a post-change parameter.
At a given iteration $Q_n(\theta)$ is obtained from a set of component functions of the form:
\begin{equation}
\label{eq:bernoulli_cost}
q^{(n, \tau)}(\theta) = a^{(n, \tau)} \log \left( \frac{\theta}{\theta_0} \right) + b^{(n, \tau)} \log\left( \frac{1-\theta}{1-\theta_0} \right), 
\end{equation}
with $a^{(n, \tau)} = \sum_{t=\tau}^n x_t$, $b^{(n, \tau)} = \sum_{t=\tau}^n (1- x_t)$. This curve represents the LR statistic for a change at $\tau$.

\begin{figure}
    \centering
    \includegraphics[width=.3\linewidth]{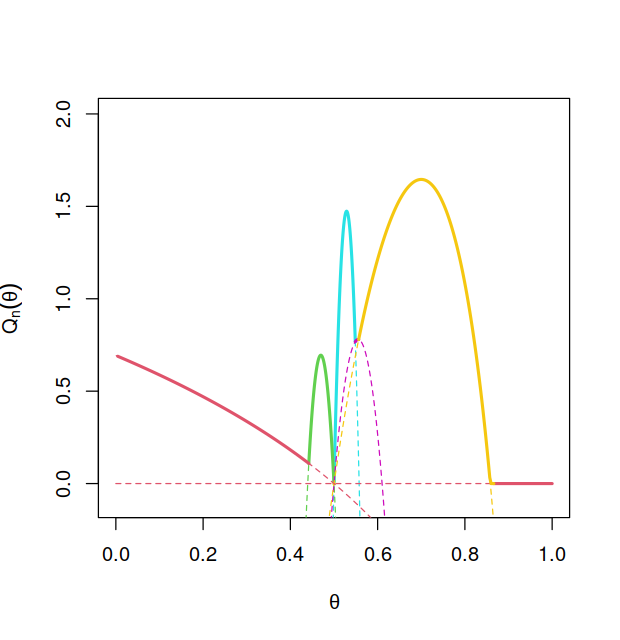}
    \caption{The components function of the functional cost $Q_n(\theta)$ at iteration $n$. The functions are plotted in different colours as dotted lines. The values for which they are optimal, e.g. they dominate all other lines and contribute to $Q_n(\theta)$, are solid. These curves are in line with those of the bottom right plot of in Fig. \ref{fig:NP-FOCuS_illustration}, illustrating the last step of our algorithm for $\theta > \theta_0 = 0.5$.}
    \label{fig:functional_cost_curves}
\end{figure}

Whilst there are $n$ such possible curves, each one relative to a different candidate changepoint $\tau$, in practice $Q_n(\theta)$ can be defined in term of a much smaller number of curves. In other words, for a large $n$, only a small proportion of these curves will contribute to the functional cost, \textit{i.e.} will be the components of our piecewise function $Q_n(\theta)$, such that for some values of $\theta$ we have that $q^{(n, \tau)}(\theta) = Q_n(\theta)$. In our example, we plot in Figure \ref{fig:functional_cost_curves} the the curves contributing to Figure \ref{fig:functional_cost_ill}. If there are no values of $\theta$ for which a curve contributes to the optimal cost, we can drop it, a mechanism that is known as \textit{pruning} that leads to the speed-ups in the algorithm. Therefore, Ber-FOCuS presents a way of iteratively  and efficiently updating $Q_n(\theta)$, figuring out which are its components' functions and finding its global maximum to solve \ref{eq:functional-recursion} exactly.

If $k$ is the number of functions that contribute to $Q_n(\theta)$, we can represent $Q_n(\theta)$ on a computer efficiently as an ordered set of $k$ triples $\mathrm{Q}_n = \{q_{i} = (a_{i}^{(n, \tau_i)},\ b_{i}^{(n, \tau_i)}, l_{i}^{(n, \tau_i)}),\ \ i = 1, \dots,\ k\}$.
Each triple represents a curve $q^{(n, \tau_i)}(\theta)$, and contains the coefficients of $\log(\theta/\theta_0)$ and of $\log( (1-\theta)/(1-\theta_0) )$, and a value $l_i$ which is the left-most value of $\theta$ for which the associated function is contributing to the optimal cost, \textit{i.e.} $\theta \in [l_{i}^{( \tau_i, n)}, l_{i+1}^{(\tau_{i+1}, n)})$, $q_{i}^{(n, \tau_i)}(\theta) = Q_n(\theta)$. In this interval our curve will be greater than all the other curves and the zero line.

For brevity, for the rest of this section, we will fix the iteration. We  can simplify the notation by dropping the superscript $(n, \tau)$, and hence we will denote our triples and functions simply as $q_1,\ldots,q_k$. This is possible as $n$ is fixed and $\tau$ is redundant information: by construction 
$\tau_i = n - (a_i + b_i)$ as we are dealing with binary data.
We will assume that the triples are 
ordered in such a way that $\theta_0= l_{1} < \dots < l_{k} < l_{k+1}=1$, so
that for all $\theta \in [l_i, l_{i+1})$,  $Q_n(\theta) = a_i \log(\theta/\theta_0) + b_i  \log( (1-\theta)/(1-\theta_0)$. 



\begin{figure*}
    \centering
    \includegraphics[width=.6\linewidth]{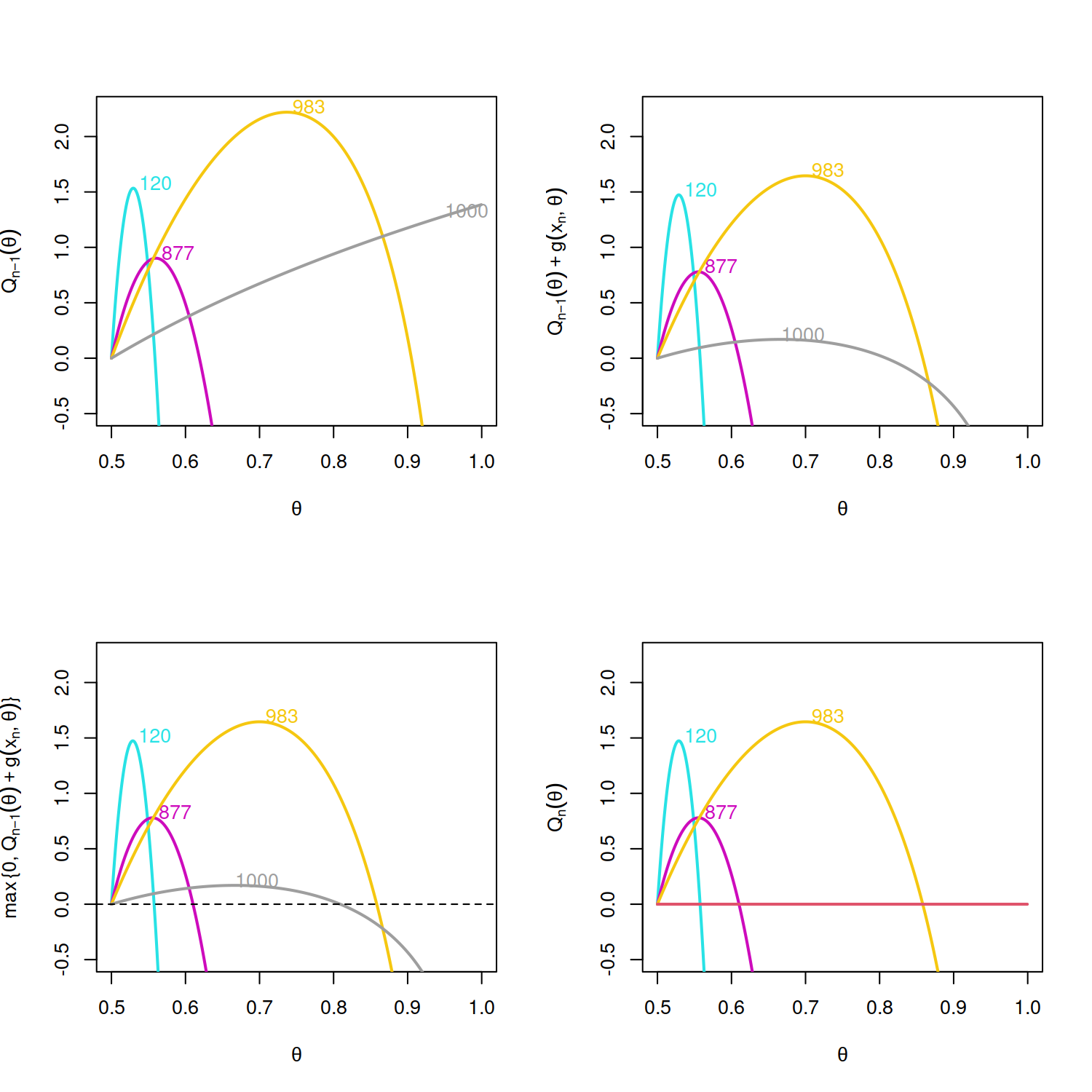}
    \caption{One iteration of the NP-FOCuS procedure illustrated, for $\theta>\theta_0$. In this example, we assume $\theta_0 = 0.5$. Top-left: the cost function from the previous iteration $Q_{n-1}(\theta)$. The total cost is given by the maximum of 4 functions: we can see how each of those is optimal for some values of $\theta$. Labels refer to the time $\tau$ at which each function was introduced. Top-right: the update step. By adding the new observation $x_n$, we update the coefficients $a_{i, \tau}$ or $b_{i, \tau}$ of each function. Although the shape of our function has changed, the points where quadratics intersect are unchanged. Bottom-left, the pruning step. Here we compare the functions with the zero line (dashed line). We find the lowest value of $\theta$ for which each function is optimal.  We notice how for the grey function (1000) there is no such value of $\theta$ for which the function is optimal, as the intersection with its closest quadratic falls below the zero line. We can therefore safely prune the function. Bottom-right: $Q_n(\theta)$, after we add a new piece (in red).}
    \label{fig:NP-FOCuS_illustration}
\end{figure*}

A description of the full Bernoulli FOCuS (Ber-FOCuS) procedure is given in Algorithm \ref{alg:NP-FOCuS_base_recursion}, with a graphical illustration of the procedure in Figure \ref{fig:NP-FOCuS_illustration}.
For each iteration, we have three main steps. Step 1, the update step, updates the values of the coefficients of stored functions according to \eqref{eq:cost_ber}. I.e., we update the count of positive or negative observations since the introduction of a curve. This is a simple increment of either coefficient $a_{i}$ or $b_{i}$. Step 2, which will be detailed in Section \ref{sec:pruning}, is the functional pruning step and focuses on identifying which pieces are no longer optimal (and will never be in the future). Finally, the optimization step, where we maximise the total cost $Q_n(\theta)$ over $\theta$, to get our test statistics. This is equivalent to taking the maximum of all maximums of our functions. As we will see in Section \ref{sec:complexity} to reconstruct the optimal cost we need only to store a small subset of candidate changepoints: at the $n^{th}$ iteration we expect to store on average $k \approx \log(n)$ functions.

\begin{algorithm}[htb]
	\caption{Ber-FOCuS (one iteration)}
	\label{alg:NP-FOCuS_base_recursion}
	\KwData{$x_n \in \{0, 1\}$ the data point at time $n$.}
	\KwIn{$Q_{n - 1} (\theta)$ the cost function from the previous iteration.}
	
	$\Tilde{Q}(\theta) \arr  Q_{n-1}(\theta) + g(x_n, \theta)$
	
	$Q_n(\theta) \arr \max \left \{0, \ \Tilde{Q}(\theta) \right\} $ \tcp*{see Algorithm \ref{alg:melkmans}}
	
	$\mathcal{Q}_n \arr \max_\theta Q_n(\theta)$
	
	\Return{$Q_n(\theta)$ for the next iteration, $\mathcal{Q}_n$ as the test statistic.}
\end{algorithm}

\subsection{Step 2: Efficient Pruning of the Bernoulli cost}\label{sec:pruning}

Functional pruning consists of restricting the set of values to consider as candidate changepoints on account of the information they carry up to time $n$. That is, at each iteration we can stop considering functions introduced at times if they will never be optimal for our cost $Q_n(\theta)$. 
To achieve this, \cite{romano2021fast} propose an efficient pruning rule, we can adapt to our setting.
As in \cite{romano2021fast}, we propose to update $Q_n(\theta)$ separately for $\theta > \theta_0$ and $\theta < \theta_0$. We will report the update rule for $\theta > \theta_0$ alone as, by symmetry, the update for $\theta < \theta_0$ can be found by applying the same algorithm  with the data inverted, i.e. replacing $x_t$ with $(1-x_t)$.

If we consider Ber-FOCuS algorithm of Algorithm \ref{alg:NP-FOCuS_base_recursion}, then at the start of an iteration we will have a set of triples. The first step, the update step, only affects the $a_i$ and $b_i$ coefficients. The $l_i$ components, which specify the intervals of $\theta$ that each curve is optimal, are unchanged. This follows as the differences between the curves are unaffected by the update, as each curve is changed by adding the same function $g(x_n,\theta)$, and it is the difference between curves that determines where one if greater than all others.

The pruning of curves only occurs in step 2, and is as a result of taking the maximum of $\tilde{Q}(\theta)$ and the zero line. The idea is that we need to find the values of $\theta$ for which the zero line is optimal. 
Define $\textsf{root}(q_{i})$ to be the largest value of $\theta>\theta_0$ such that $q_{i}(\theta) = 0$, if such a value exists. 
It is straightforward to show that, for a given function $q_{i}$ that contributes to $\tilde{Q}(\theta)$ for $\theta>\theta_0$, if $\textsf{root}(q_{i})<1$ exists, then the zero-line is better on the interval: 
\[
\left[\textsf{root}(q_{i}), \ 1 \right).
\]
Considering all functions, this implies that  if
$\max_{\tau} \ \textsf{root}(q_{i}) <1$ then the zero line is globally optimal on $[l^*,1)$, where
\[
l^* = \max_{\tau} \ \textsf{root}(q_{i}).
\]
Therefore, the triple defining the zero line will be given by $(0, 0, l^*)$. Given that the quadratics are ordered, it means that any quadratic with $l_i <  l^* $ can therefore be removed. Finally, the ordering of the quadratics means that $l_i < l_j$ if and only if $i<j$. For this reason, we can start checking from the last quadratic $q_{k}$ and stop as soon as the pruning condition is not met.
The pruning procedure is summarised in Algorithm \ref{alg:melkmans}. 
Following the update, for pruning, given that at any point in time $l_i = \textsf{root}(q_{i} - q_{i-1}) \leq l^*$ and that $q_i(\theta_0)=0$, we know that whether $q_i(l_i)<0$ there will be no such value of $\theta$ for which $q_i$ will be optimal.
In fact, we know that $q_i$ will be greater than $q_{i-1}$ only starting from $l_i$, and if $q_i(l_i) < 0$ falls below the zero line, then $ q_{i-1}(\theta) < q_i(\theta) < 0 \ \forall \ \theta \in (l_i, 1)$ and therefore $q_i$ will be never be optimal and can be safely pruned. The bottom-left plot of Figure \ref{fig:NP-FOCuS_illustration} can be of aid in understanding the pruning and the following theorems.
  

\begin{algorithm}[htb]
	\caption{Algorithm for $ \max \{0, \ \Tilde{Q}(\theta) \} $ for $\theta>\theta_0$}
	\label{alg:melkmans}
	\KwIn{
 $\Tilde{Q}$, an ordered set of triples $\{q_i = (a_i,\ b_i, l_i)\ \forall \ i = 1, \dots,\ k\}$}
 
	$i \arr k$;\\
	\While {$ q_i(l_i) \leq 0$ \mbox{and} $i \geq 1$}{
		
		$i \arr i - 1$;\\
	}
	
	\If{$i \neq k$} {
		$\mathrm{Q} \arr \mathrm{Q} \setminus \{q_{i+1}, \ \dots, q_{k}\}$\tcp*{drop pieces}
		$k \arr i$
	}

    \uIf{k = 0} {
        $q_{k+1} \arr (0, 0, \theta_0)  $\tcp*{add first piece}
	} 
    \Else {
    $q_{k+1} \arr (0, 0, \textsf{root}(q_k))  $\tcp*{add new piece}        
    }
	$\mathrm{Q} \arr \{\mathrm{Q}, q_{k+1}\}$

	\Return{$\mathrm{Q}$}
\end{algorithm}

We therefore learn that the algebraically ideal version of the algorithm relies on computing the value $\textsf{root}(q_{k})$ in step 9 at each iteration. However, given the shape of our cost function, we do not have a closed form for computing this root. A naive implementation would thus require obtaining such a value numerically. 


Fortunately, there is an alternative way of performing the same pruning exactly that avoids computing the root value. As we will learn in the following Theorem and Lemma, this will be achieved by studying the link between the value of the roots of our pieces and their argmax. 

\begin{theorem}
\label{theo:pruning}
    Assume $q_i$ and $q_j$ are two curves that contribute to $Q_n(\theta)$ for $\theta>\theta_0$, with $i>j$.
    Then, $\textsf{root}(q_{i}) < \textsf{root}(q_{j})$ if and only if $ a_i/(a_i + b_i) < a_j/(a_j + b_j)$.
\end{theorem}

\noindent {\bf Proof.} We begin with some properties of the curves. A curve $q_i$ has $\arg\max q_i(\theta) = a_i/(a_i + b_i)$. By construction, if $i > j$ then $a_i \leq a_j$ and $b_i \leq b_j$. Finally $q_i(\theta_0) = q_j(\theta_0) = 0$. 

We now link the $\arg\max$ of our functions to their roots. 
First we can rescale each curve without changing the location of its maximum of its root. For curve $q_i$ define $\Tilde{a}_i = a_i/(a_i+b_i), \ 1 - \Tilde{a_i} = b_i/(a_i+b_i)$ and write
\begin{equation}
    \Tilde{q}_i(\theta) = \Tilde{a} \log \left( \frac{\theta}{\theta_0} \right) + (1-\Tilde{a}) \log \left( \frac{1-\theta}{1-\theta_0} \right).
\end{equation}
We then have that
\begin{equation}
    \textsf{root}(\tilde{q}_i) \geq \textsf{root}(\tilde{q}_j) \iff \Tilde{a}_i \geq \Tilde{a}_j,
\end{equation}
This follows by noting that $\Tilde{q}_i(\theta)$ is unimodal as $\frac{\partial \Tilde{q}_i(\theta)}{\partial \theta}$ has one zero, and that $\forall \ \theta \in [0, 1], \frac{\partial \Tilde{q}_i(\theta)}{\partial \theta} \geq \frac{\partial \Tilde{q}_j(\theta)}{\partial \theta} \iff \Tilde{a}_i \geq \Tilde{a}_i$.

As $q_i$ has the same root as $\tilde{q}_i$, the result follows immediately.
\hfill $\Box$
\vspace{5pt}


We will use this result to simplify the pruning procedure in the following Lemma. 
\begin{lemma}
    The condition for pruning $q_i(l_i) \leq 0$ at step 2 of Algorithm \ref{alg:melkmans} is implied by $a_i/(a_i + b_i) < a_{i - 1}/(a_{i - 1} + b_{i - 1})$ for $i > 1$ and $a_i/(a_i + b_i)<\theta_0$ for $i=1$.
\end{lemma}

\noindent {\bf Proof.}
To link the pruning condition and the results from Theorem \ref{theo:pruning} we need to show that for $i > 1$ whether $q_i(l_i) \leq 0$ we find $\textsf{root}(q_{i}) < \textsf{root}(q_{i-1})$.
We start by noting that $l_i$ is the root of the function $q^*(\cdot) = q_i(\cdot) - q_{i-1}(\cdot)$, i.e. $l_i = \textsf{root}(q^*)$, and that $q^*(\theta) \geq 0$ on $[\theta_0, l_i]$.

Then, if $q_i(l_i) \leq 0$ we have
$$
q_{i-1}(\theta) = q_{i}(\theta) + q^*(\theta) > q_i(\theta) \ \text{for} \ \theta \in (\theta_0, l_i),
$$
and as $\textsf{root}(q_{i-1}) \in (\theta_0, l_i)$, it follows that 
$$
q_i(\textsf{root}(q_{i-1})) < q_{i-1}(\textsf{root}(q_{i-1})) = 0.
$$
By concavity of $q_i$ then $\textsf{root}(q_{i}) < \textsf{root}(q_{i-1})$.
The argument for $i = 1$ follows simply by noting that in this case $l_1 = \textsf{root}(q_1) < \theta_0 \iff a_1/(a_1 + b_1)<\theta_0$.
\hfill $\Box$
\vspace{5pt}

Hence we can swap the pruning condition involving the root-finding with the simpler condition: $a_i/(a_i + b_i) < \max \{ \theta_0, \ a_{i}/(a_{i} + b_{i})\}$. Given that this novel pruning only involves the $a_i, b_i$ values, it is no longer necessary to find and store the $l_i$ value at each iteration. { This result relies, in part, on the functions, $q_i(.)$, being unimodal. Similar results apply for other functions, and these have been used to generalise the FOCuS algorithm to other one-parameter exponential family models \cite[see][]{ward2023constant}.} 



\subsection{Extension to the $\theta_0$ Unknown Case}

We next extend our recursion to the case where both $\theta_0$ and $\theta_1$ are unknown. Assume we are observing a stream of Bernoulli random variables $x_1, \dots, x_n$ distributed as a $Ber(\theta_0)$ under the null and as a $Ber(\theta_1)$ under the alternative. 
The likelihood ratio test statistic is
\begin{align}
\mathcal{Q}_n &= -\max_{\theta \in \mathbb{R}} \sum_{t=1}^n h(x_t, \theta) +\\
&+\underset{\substack{\tau \in \{ 1, \dots, n-1 \} \\ \theta_0, \theta_1 \in \mathbb{R}} }{\max} \left\{\sum_{t=1}^{\tau} h(x_t, \theta_0) + \sum_{t=\tau+1}^n h(x_t, \theta_1) \right\},\label{eq:LR-theta0theta1}
\end{align}
where
\[
h(x_t, \theta) = x_t \log\theta + (1-x_t) \log(1-\theta).
\]

Solving this directly for all possible values of $\tau$, via directly storing the partial sums, will result in a procedure that has computational complexity of $\mathcal{O}(n)$ per iteration and $\mathcal{O}(n)$ in memory.
We employ a functional pruning approach and solve \eqref{eq:LR-theta0theta1} exactly through the following recursion. 
\begin{proposition}
	\label{theo:propositiontheta0}
	Let $Q_0(\theta) = 0$ and for $n = 1, 2, \dots$ let $Q_n$ be defined by the recursion:
	\[
	Q_n(\theta) = \max \left\{ \max_{\theta} \sum_{t = 1}^{n} h(x_t, \theta), \ {Q}_{n - 1}(\theta) + h(x_n, \theta) \right\}.
	\]
	Then, at time $n$, $\mathcal{Q}_n = \max_{\theta} \sum_{t = 1}^n h(x_t, \theta) - \max_\theta Q_n(\theta).$
\end{proposition}
The proof to Proposition \ref{theo:propositiontheta0} is found in Appendix \ref{app:proofs-proposition}. 

This recursion can be solved using the same ideas as from \cite{romano2021detecting}. As before $Q_n(\theta)$ will be the maximum of a set of curves. Each curve will be of the form
\[
a\log(\theta)+b\log(1-\theta)+c,
\]
where the coefficients will depend on the changepoint location associated with that curve. Importantly the set of changepoint locations that we need to store curves correspond to locations whose curves would be stored for the $\theta_0$ known case for some $\theta_0$. The only difference that $\theta_0$ has on the changepoint locations whose curves are kept is that when considering positive changes we only need consider $\theta_1>\theta_0$, and for negative changes we need only consider $\theta_1<\theta_0$. As we need to keep curves that are optimal for some value of $\theta_0$ this means that for positive changes we need to prune the same curves as for the $\theta_0$ known case by for $\theta_0\rightarrow 0$. For negative changes we prunes as for the $\theta_0$ known case but for $\theta_0 \rightarrow 1$. 

The algorithm to implement Ber-FOCuS in the $\theta_0$ unknown for $\theta_1>\theta_0$ is almost identical to Algorithm \ref{alg:melkmans} except  we now store a 4-tuple $(a_i,b_i,c_i, l_i)$ for each curve, so in step 9 the new curve is described by the 4-tuple $(0,0,c, \textsf{root}(q_k))$  with $c=\max_\theta \sum_{t = 1}^n h(x_t, \theta)$. As before, in practice we can implement the resulting algorithm without storing $l_i$.

\subsection{Aggregating the Bernoulli Traces and NP-FOCuS}\label{sec:aggregators}

Having covered how to independently monitor the $M$ Bernoulli processes for all quantiles, we describe how to obtain a global statistic for NP-FOCuS. 

Our approach is to consider two aggregation functions. One is to take the maximum of the statistics, and the other is to take the sum, as in e.g., \citet{mei2010efficient}.  

{\bf NP-FOCuS:}
	Let $Q^{1}_n(\theta), \dots, Q^{M}_n(\theta)$ be the costs at time $n$ for the $M$ Bernoulli sequences $\mathbb{I}(y_n \leq p)\, p \in \{p_1, \dots, p_M\}$, and let the Ber-FOCuS statistics:
	\[
	\mathcal{Q}^{m}_n = \max_\theta Q^m_n(\theta)
	\]
	Then, we will detect a changepoint at time n whether: 
	\[
	\sum_{m = 1}^M \mathcal{Q}^{m}_n \geq \xi^{sum} \ \text{or} \ \max_{m \in \{1, \dots, M\}} \mathcal{Q}^{m}_n \geq \xi^{max},
	\]
	with $\xi^{sum},\ \xi^{max} \in \mathbb{R}$.
A formal description of the NP-FOCuS algorithm is reported in Algorithm \ref{alg:NP-FOCuS}.

\begin{algorithm}[htb]
	\caption{NP-FOCuS (one iteration)}
	\label{alg:NP-FOCuS}
	\KwData{$y_n \in \mathbb{R}$ the data point at time $n$.}
	\KwIn{$\xi^{sum},\ \xi^{max}, \{p_1, \dots, p_M\}$, $\{Q^{1}_{n-1}(\theta), \dots, Q^{M}_{n-1}(\theta)\}$.}
	
	\For {$m = 1, \dots, M$}{
		$Q^m_{n}(\theta), \mathcal{Q}^m_n \arr \text{Ber-FOCuS}(x_n = \mathbb{I}(y_n < p_m);\ Q^m_{n-1}(\theta))$\tcp*{Algorithm \ref{alg:NP-FOCuS_base_recursion}}
	}
	$\mathcal{S}^{sum} \arr \sum_{m = 1}^M \mathcal{Q}^{m}_n$;\\
	$\mathcal{S}^{max} \arr \max_{m \in \{1, \dots, M\}}$;\\
	\If{$\mathcal{S}^{sum}_n \geq \xi^{sum}\ or \ \mathcal{S}^{max}_n \geq \xi^{max}$}{
		\Return{$n$ as a stopping point}.
	}
	
	\Return{$\{Q^{1}_{n}(\theta), \dots, Q^{M}_{n}(\theta)\}$ for the next iteration.}
\end{algorithm}

One drawback of having to monitor two streams from two different aggregators comes with the procedure requiring two thresholds. However, there is an advantage in monitoring both streams simultaneously, as there are change scenarios where we expect one to perform better than the other. The sum statistic should have greater power to detect small shifts in the distribution that affect all or many quantiles. The maximum statistic will have greater power to detect larger shifts that affect e.g. just the tail of the distribution. 
By setting either $\xi^{sum}$ or $\xi^{max}$ to infinity, our method would result to considering a test just based on the maximum or the sum. 

\subsection{Tuning Strategy and Choice of Quantiles}\label{sec:tuning}

We now describe how to tune the initial parameters of the NP-FOCuS procedure. Quantile values can be computed either on training observations or on a probation period, in case of no training data.
Rather than taking evenly spaced quantiles, we build geometrically spaced quantiles following the approach of \cite{Haynes}. That is, for fixed $M$, we take $p_1, \dots, p_M$ in such a way that $p_m$ is the empirical quantile with probability
\[
\left\{ 1 + (2 n - 1) \exp \left[ - \frac{(2 m - 1)}{M} \log(2 n - 1) \right] \right\}^{-1}.
\]

This is to give more importance to values in the tail of the distribution of the data. In case of a change in the tails of a distribution, for instance, this would allow for a quicker detection as the traces of the more-extreme quantiles are more likely to pass the threshold for the max statistic.

\begin{figure}
    \centering
    \includegraphics[width = .38\textwidth]{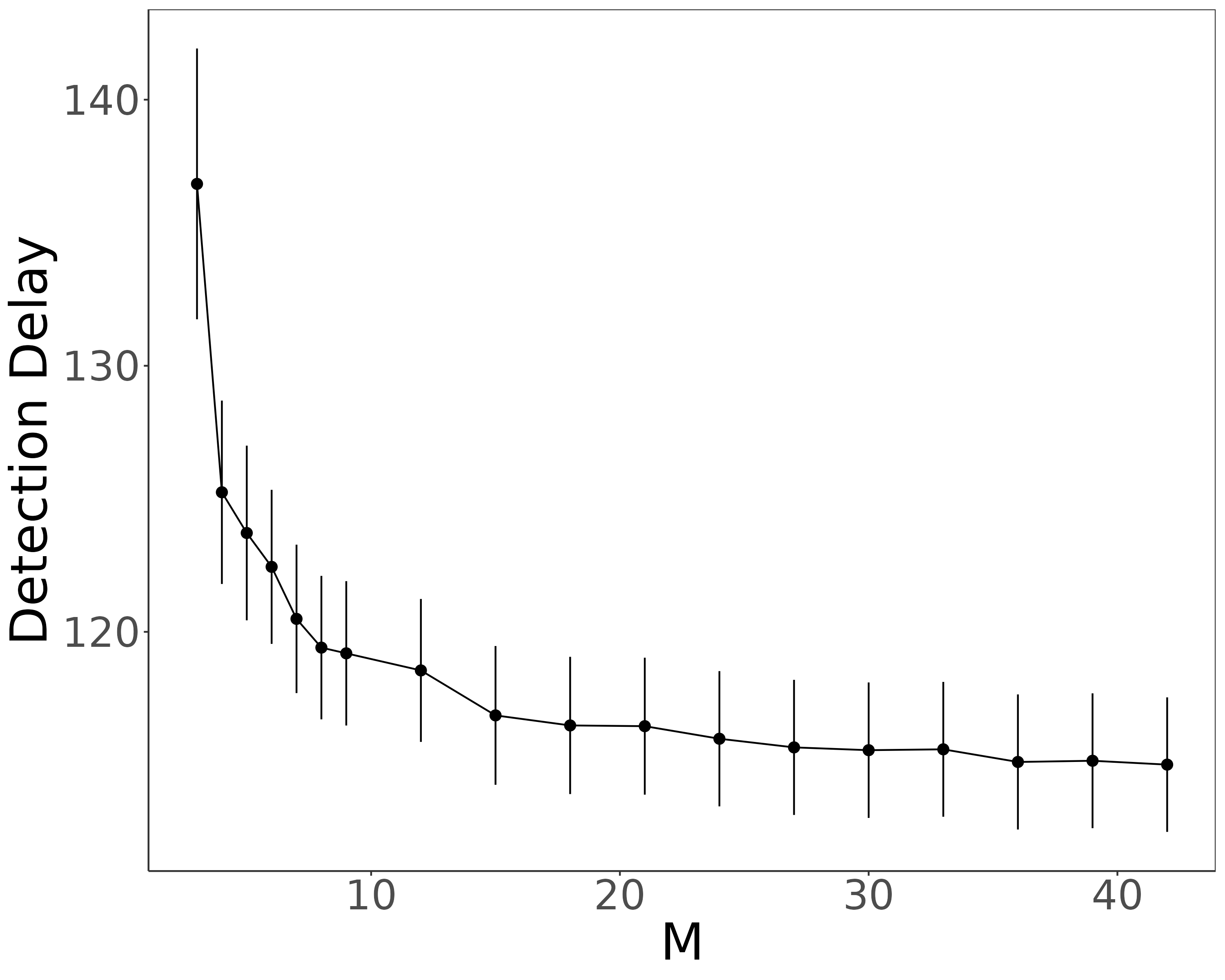}
    \caption{Average Detection Delay in function of the number of quantiles M. Vertical bars represent the standard deviation across the various replicates.}
    \label{fig:choice_of_quantiles}
\end{figure}

As we will find out from Section \ref{sec:complexity}, the choice of $M$ affects the computational complexity of the procedure. Even if a higher number of quantiles should lead to better statistical power, in practice there is not much of a gain in picking values of $M$ greater than 15. This can be checked from the elbow plot in Figure \ref{fig:choice_of_quantiles}, where we measure the average detection delay as a function of $M$. Simulations were performed as described in Section \ref{sec:simulations}. 

Finally, we tune the thresholds  $\xi^{sum}$ or $\xi^{max}$ via a Monte Carlo approach. This is common practice in the literature when using fixed thresholds. In particular, we follow the approach of \citet[Section 4.1]{chen2022high}.
We generate, say, $T$ many sequences as long as the desired average run length $N$ under the null. This can be achieved either by simulating the process directly or bootstrapping from training data. We then compute both the sum and the max statistics on all the $T$ sequences and pick the max on each. The idea is that the stopping times $S^{sum}, S^{max}$, the times in which respectively the sum or the max statistics go over the thresholds $\xi^{sum},\ \xi^{max}$, are approximately exponentially distributed. This means that we want to choose the thresholds so that the probability of stopping after time $N$, if there is no change, is {$\exp(-1)$}. 

Based on our simulations we  pick the estimated thresholds $\tilde{\xi}^{sum},\ \tilde{\xi}^{max}$ to be the smallest values such that the proportion of times we do not detect a change by time $N$, based on the sum and max statistics respectively, is less than { $\exp(-1)$}. This approach would give thresholds with the approximate average run length we require if we were using only the max or only the sum statistic. As we are using both, we now fix the ratio of $\tilde{\xi}^{sum}$ to  $\tilde{\xi}^{max}$ but use the same procedure to scale their values to get a composite test with the correct average run length.


\section{Computational Complexity of the NP-FOCuS Procedure}\label{sec:complexity}

The computational and memory complexity per iteration of Algorithm \ref{alg:NP-FOCuS} is $M$ times the cost of running Ber-FOCuS. The computational cost of Ber-FOCuS consists of the pruning step and the update and maximisation steps. 

The pruning step has a computational cost that is, on average $O(1)$ per iteration. This is because at each iteration we will consider some number, $m$ say of curves -- and this will result in at least $m-1$ curves being removed. As we only add one curve per iteration, and a curve can only be removed once, this limits the average number of curves to be considered per iteration to be at most 2.

The other steps have a computational cost that is proportional to the number of curves that are kept. Furthermore, empirically, these step dominate the computational cost of the algorithm. We can bound the expected cost of these steps with the following result which bounds the number of curves that are stored.

\begin{theorem}\label{theo:funcbound}
Let $x_1, \dots, x_n, \dots$ be a realization of an independent Bernoulli process centred on $\theta_0$. Let the number of functions stored by Ber-FOCuS for $\theta>\theta_0$ at iteration $n$ be 
$\#\mathcal{I}_{1:n}$. Then if there is no change prior to $n$
\[
E(\#\mathcal{I}_{1:n}) \leq (\log(n)+1),
\]
while if there is one change prior to $n$ then
\[
E(\#\mathcal{I}_{1:n}) \leq 2(\log(n/2)+1).
\]
\end{theorem}
The proof of this theorem is found in Appendix \ref{app:proofs-theo3}. By symmetry, the theorem extends to values of $\theta < \theta_0$. From the theorem, we learn that, at each iteration, we expect to check $k = \log(n) + 1$ curves in case a change has not occurred, and $k = 2 \log(n/2) + 1$ curves in the case where a change has been occurred but is undetected.


    

\section{Empirical Evaluation of the Method}

\subsection{Simulation study}\label{sec:simulations}

\begin{figure}[tb]
    \centering
    \includegraphics[width=\linewidth]{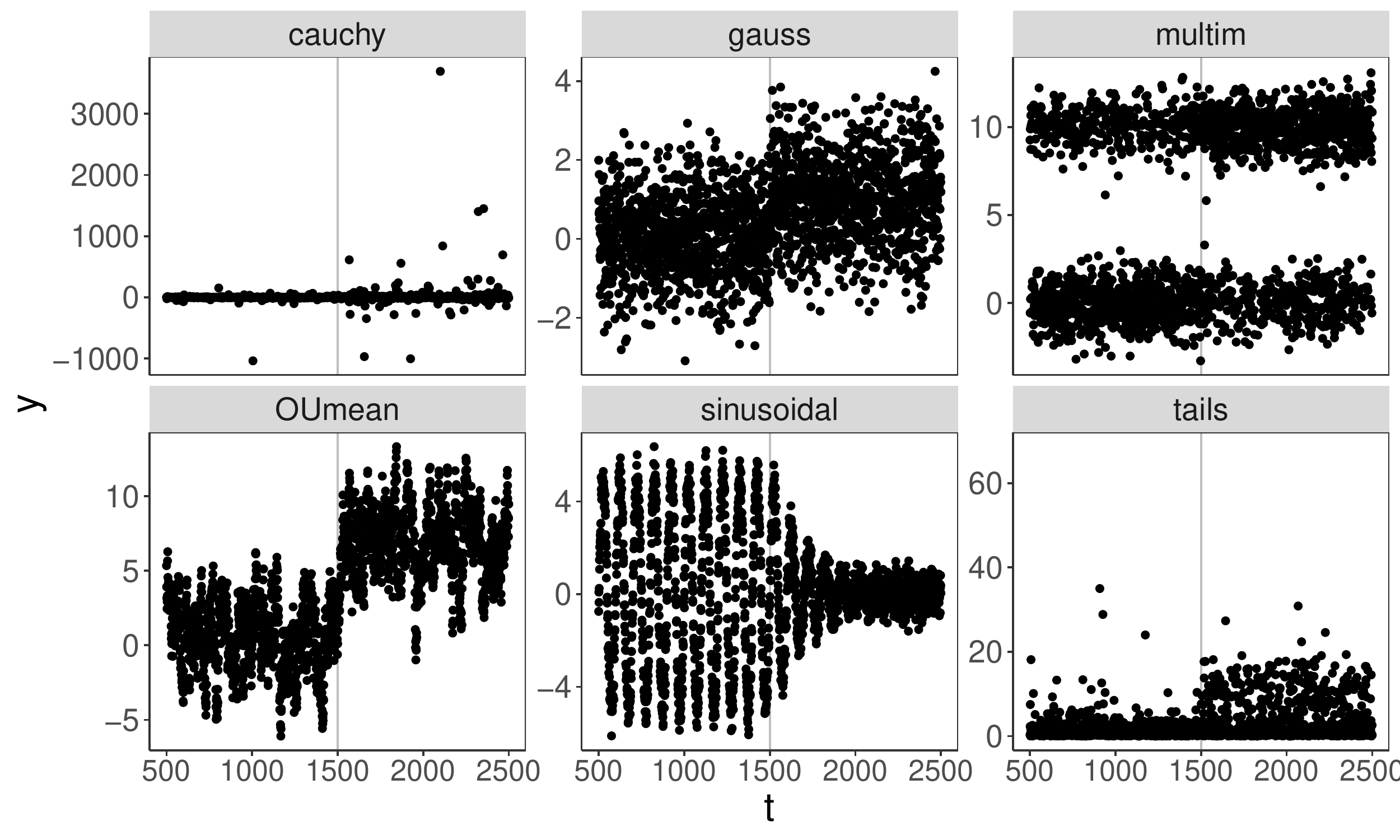}
    \caption{Example sequences from the six different scenarios considered for the simulation study. Solid grey lines demarcate a changepoint.}
    \label{fig:sim_scenarios}
\end{figure}

We performed a simulation study to assess performances of $\NPF{}$ and other online changepoint procedures in a variety of scenarios, illustrated in Figure \ref{fig:sim_scenarios}. Those scenarios were chosen to benchmark procedures over a range of different challenges in online changepoint detection. Specifically, present scenarios for Cauchy change-in-scale, Gaussian change-in-mean, change-in-mode in a mixture of two Gaussian distributions, change-in-mean in an Ornstein-Uhlenbeck process, decay in a sinusoidal process with random noise, and a change-in-tails scenario -- we cover in details how to generate these in the Supplementary Materials, Section \ref{sec:sim_descr}. All the scenarios show \textit{i.i.d} sequences with the exception of the OU process and the Sinusoidal process. These two were added to account for changes in scale and location in presence of strong temporal dependency, a well-known challenge in changepoint detection which is present in many real-world applications \citep[see][]{romano2021fast, cho2020multiple, hallgren2021changepoint}.

We compare $\NPF{}$ with the online non-parametric procedure NUNC \citep{austin2023online}, the method from \cite{ross2021nonparametric}, the NEWMA procedure with RFF \citep[see Section IV of][]{kerivenNEWMA} { and the recent online kernel CUSUM procedure from \cite{wei2022online} with Gaussian Kernel (that we call Wei-CUSUM). 
As with other methods, { Wei-CUSUM} procedure assumes independent data. However, whereas the other methods can be applied to non-independent data by increasing the threshold for detecting a change, this is impossible for such a procedure. In fact, in line with other Kernel based online statistics, such as KCUSUM from \cite{flynn2019change}, if the data is serially dependent then the contribution to the statistic for each new data point will on average by positive even if there is no change -- and thus the method will be prone to false positives regardless of how large the threshold for the test is chosen.
Thus 
we exclude Wei-CUSUM from the sinusoidal and OU scenarios.} Lastly we add to the comparison the FOCuS procedure for Gaussian change-in-mean from \cite{romano2021fast}. A robust bi-weight loss was employed in this case to account for the presence of outliers in some scenarios. 

Each experiment consisted of 100 replicates. Initial parameters were tuned according to Section \ref{sec:tuning}. Thresholds were selected to achieve an average run length of $10,000$ observations. Other parameters -- such as quantiles, the value of the bi-weight loss parameter or the NEWMA bandwidth -- were obtained over a probation period consisting of the first $100$ observations { (in Appendix \ref{sec:add_empirical} we study the effect of the length of the probation on NP-FOCuS performances).}
The NEWMA method involves monitoring the difference of two exponentially weighted moving averages of features of the data, and it can take a substantial number of observations for this difference to stabilise. To account for this we only start monitoring the NEWMA statistic after a burn-in period of 1000 observations to avoid false-positives whilst its statistics stabilise. { For Wei-CUSUM, we gave oracle knowledge of the pre-change distribution, and tuned the scale parameter of the Gaussian kernel based on the variance of the pre-change distribution.}

\begin{figure}[tb]
    \centering
    \includegraphics[width=\linewidth]{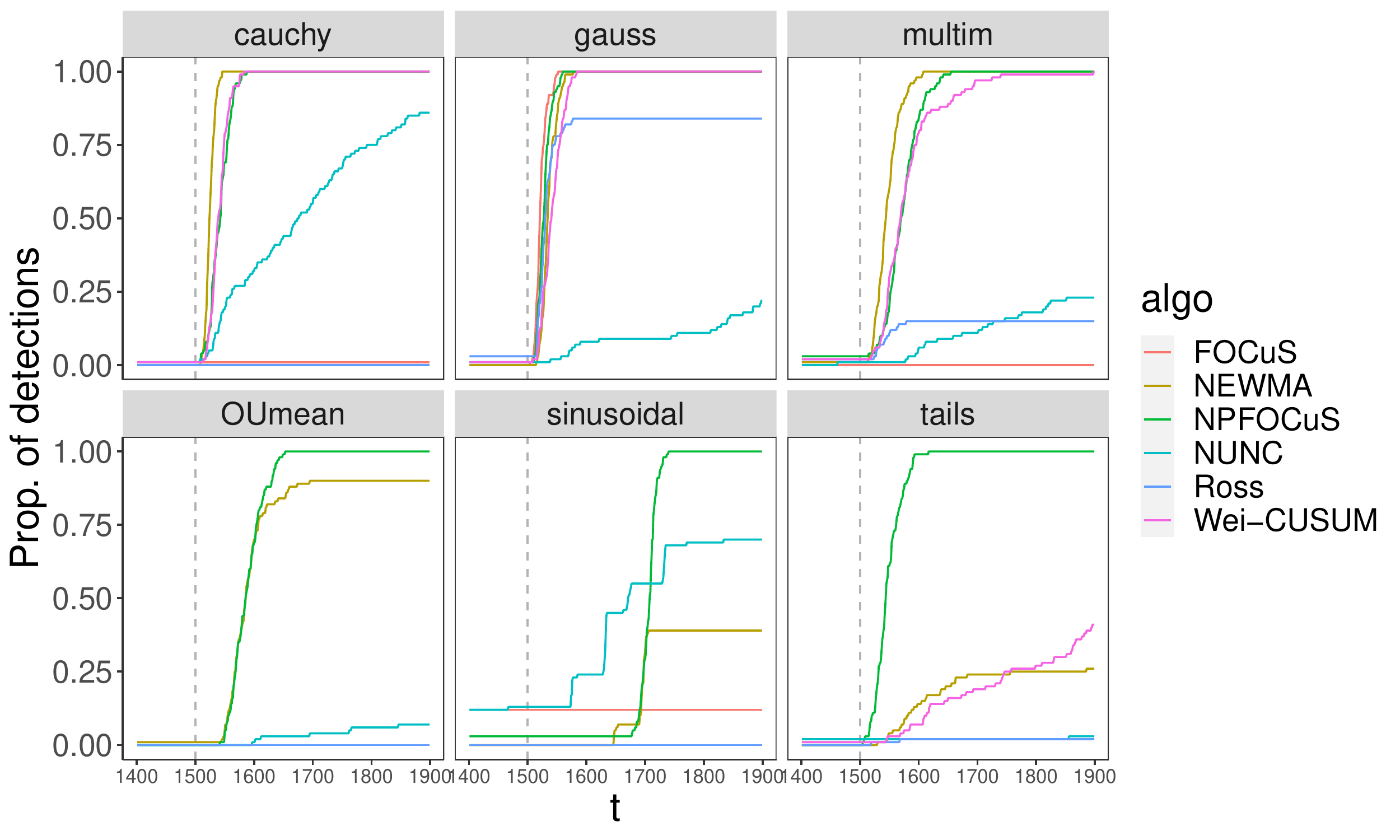}
    \caption{Proportions of changes detected within $t$ observations following the change in six different change scenarios. The change is denoted by the vertical dotted line at $t = 1000$.}
    \label{fig:prop-of-detections}
\end{figure}

We measure performances in terms of detection delay on sequences generated to have a change at $\tau = 1500$.
In Figure \ref{fig:prop-of-detections}, we report the proportions of experiments where a change was detected by time step $t$. Prior to the change, the lines show the proportion of false positives, while following the change, they show the number of true detections by a given time step: the perfect online procedure would achieve a detection within 1 observation from the change in all sequences. { Additionally, we report results in terms of average detection delay and false positive rate in Tables \ref{tab:detection_delay} and \ref{tab:fp_rate}.}

\begin{table*}[ht]
\centering
\begin{subtable}{\textwidth}
\centering
\begin{tabular}{lrrrrrr}
  \hline
  scenario & FOCuS & Wei-CUSUM & NEWMA & NPFOCuS & NUNC & Ross \\ 
  \hline
   OUmean & $>$1000 &  & 225.92 & \textbf{87.99} & 995.35 & $>$1000 \\ 
   cauchy & $>$1000 & 35.03 & \textbf{24.52} & 33.98 & 242.72 & $>$1000 \\ 
   gauss & \textbf{22.08} & 27.2 & 35.71 & 22.26 & 680.64 & 238.87 \\ 
   multim & $>$1000 & 65.44 & \textbf{42.54} & 44.86 & 943.52 & $>$1000 \\ 
   sinusoidal & $>$1000 &   & 988.15 & \textbf{165.8} & 423.91 & $>$1000 \\ 
   tails & $>$1000 & 745.03 & $>$1000 & \textbf{46.97} & $>$1000 & $>$1000 \\ 
   \hline
\end{tabular}
\caption{Average detection delay. In bold, the best result by row.}
\label{tab:detection_delay}
\end{subtable}

\vspace*{1 cm} 

\begin{subtable}{\textwidth}
\centering
\begin{tabular}{lrrrrrr}
  \hline
  scenario & FOCuS & Wei-CUSUM & NEWMA & NPFOCuS & NUNC & Ross \\ 
  \hline
 OUmean & 0.00 &   & 0.01 & 0.00 & 0.00 & 0.00 \\ 
   cauchy & 0.01 & 0.01 & 0.00 & 0.01 & 0.01 & 0.00 \\ 
   gauss & 0.00 & 0.01 & 0.00 & 0.01 & 0.01 & 0.03 \\ 
   multim & 0.00 & 0.02 & 0.01 & 0.03 & 0.01 & 0.02 \\ 
   sinusoidal & 0.12 &   & 0.00 & 0.03 & 0.13 & 0.00 \\ 
   tails & 0.02 & 0.01 & 0.00 & 0.00 & 0.02 & 0.00 \\ 
   \hline
\end{tabular}
\caption{False positive rate.}
\label{tab:fp_rate}
\end{subtable}
\caption{Average detection delay and false positive rate across the various algorithms for all scenarios of our simulation study.}
\end{table*}


We learn that, overall NP-FOCuS shows good performances in terms of statistical power. In the Gaussian case, unsurprisingly, the simple FOCuS with Gaussian loss has best performances overall, immediately followed by NP-FOCuS. Additionally, we found that NEWMA achieves faster detections both in the multimodal scenario, and in the Cauchy scenario, { immediately followed by NP-FOCuS and Wei-CUSUM with similar performances}. However, upon further testing, we noticed that the performance of NEWMA tends to degrade if the change occurs earlier, even if we use a shorter burn-in period (see Appendix \ref{sec:add_empirical} for further details). { Lastly, Kernel based procedures such as Wei-CUSUM,} in addition to requiring the ability to sample from the pre-change distribution, are sensitive to the choice of the kernel.

We find NP-FOCuS to be more robust to strong dependence in the signal than other procedures. The most challenging scenario for NP-FOCuS is the sinusoidal scenario, where we start to consistently estimate the change with a delay of 200 observations. This, as it can be seen from Figure \ref{fig:sim_scenarios}, corresponds to the point where there is a clear difference in scale. In this scenario, for comparison, other methods are either prone to false positives, or missed detections.



\subsection{Monitoring Power Attenuation on Optical Lines}\label{sec:application}

\begin{figure*}[tb]
    \centering
    \includegraphics[width=.7\linewidth]{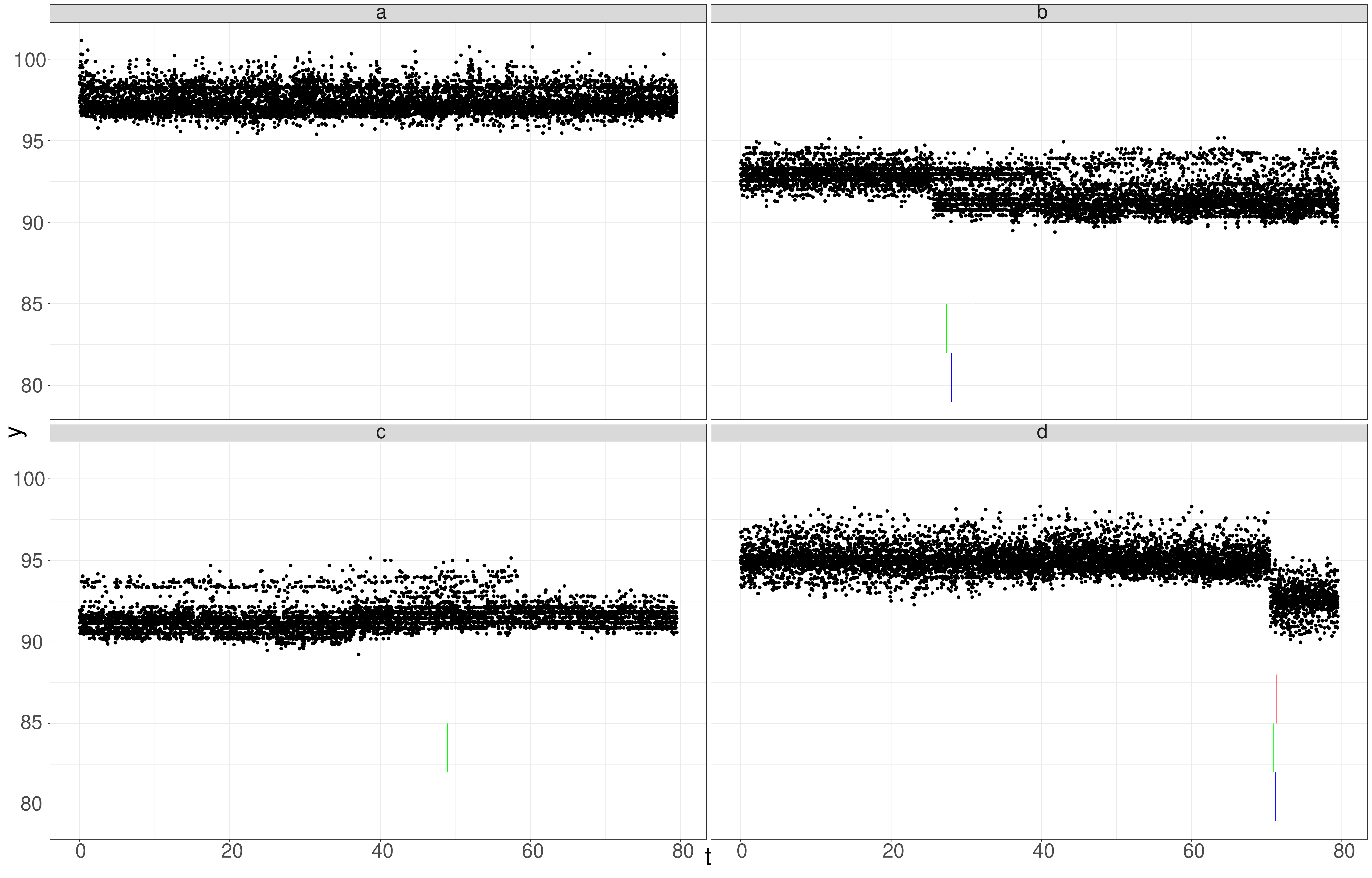}
    \caption{Four examples of the measured power attenuation over time. Red segments at the bottom of the sequences point to detection times of FOCuS for Gaussian change-in-mean (red), the presented procedure NP-FOCuS (green) and finally NUNC (blue). Respectively, we show, in a, a sequence with no change, in b, a sequence with a shift in some modes, in c, a sequence with changes in the behaviours of the tails, and finally in d an abrupt change-in-location.} 
    
    \label{fig:application_example}
\end{figure*}

We now evaluate NP-FOCuS on a real-world application, that of monitoring power attenuation on an optical communication line, a metric that is associated with the operational performance of the line. We present in Figure \ref{fig:application_example} four examples of such a metric. Under normal circumstances, the power attenuation should show a stationary behaviour over time, as seen in \ref{fig:application_example}a. Any change in distribution might therefore be of interest to the engineers and be an object of further investigation.
Efficient techniques are required as there is a need of monitoring a large number of instances over time. 


Each time-series has recordings over 80 days. The time-series were first classified by a domain expert into stationary series, and series with a change that would want to be flagged to an engineer. Thresholds were tuned over 85 non-changing sequences to achieve a false positive rate of 0.01, as described in Section \ref{sec:tuning}, with quantiles being trained over a probation period of 2 weeks in each sequence.

We draw a comparison of NP-FOCuS stopping times with both Gaussian FOCuS and NUNC. Threshold and other tuning parameters were adjusted on the same NP-FOCuS training instances to achieve comparable false positive rates. 
All 3 algorithms do not report a detection on the example without a change. In the sequences with a change, NP-FOCuS reported faster detection times. Overall, Gaussian FOCuS showed the slowest detection delay, even for the example where the change appears to be a change-in-location. This is attributable to the fact that the training data is heavy-tailed, and as mentioned in the example in the introduction this leads to slower detections caused by over-inflated thresholds. Comparing the nonparametric online procedures, NP-FOCuS and NUNC, they show similar detection times in the b and d sequences, with NP-FOCuS resulting in slightly faster detections. However, in line with the previous numerical study, NUNC struggles in detecting changes in the tails of a distribution. 

\section{Remarks and Conclusions}

We presented a nonparametric approach for online changepoint detection. The procedure keeps track of the eCDF of a data stream over a set of fixed quantiles. Keeping track of which observations are above or below each quantile maps the problem into a Bernoulli change-in-rate problem. We derived a  functional pruning sequential LR test to perform such a task efficiently in almost linear time. 



While we have provided theoretical guarantees about the computational aspects of our procedure, we have not given any results on the average run length or the expected detection delay. 
Obtaining tight results is challenging for our approach due to additional complexity of allowing for any post change (and potentially pre-change) parameter, and due to combining of information across different quantiles. For theoretical results for the related off-line changepoint procedure, including showing that in the limit where we increase the number of quantiles, we are able to have power to detect any change in distribution, see \cite{zou2014nonparametric}.

The only assumption required by the NP-FOCuS, that of i.i.d. observations, can lead to a drop in performances when dealing with time-dependent sequences. In presence of strong dependence, to achieve comparable run lengths to the i.i.d. case, thresholds are to be inflated, resulting in a slower detection delay. An approach that can account for the time dependence, such as those of \cite{romano2021detecting,cho2020multiple}, could be an avenue for further research.
Relying on fixed quantiles also means that the procedure needs some training observations to understand the range of the data. This can be a problem in particular when restarting the procedure. If the quantiles are not known, in fact, those are needed to be re-estimated at every detected change. This means that no detection will be made for the entire duration of the probation period. One simple solution would be to store a rolling window of observations of the length of the probation period. Then, as soon as a detection is made, the procedure could be restarted with quantiles computed over those stored observations.

\section{Acknowledgments}

The authors would like to thank Trevor Burbridge at BT for the helpful conversations that helped motivate the presented procedure. We are grateful to the Associate Editor and reviewers for their thoughtful comments and suggestions. The authors acknowledge the financial support of the EPSRC and Lancaster University grants EP/N031938/1, EP/R004935/1 and Mathematical Sciences RA (EP/W522612/1).

\bibliographystyle{apalike}
\bibliography{bibliography}

\newpage

\appendix

\setcounter{page}{1}

\subsection{Proofs} \label{app:proofs}

\subsection*{Proof of Proposition \ref{theo:propositiontheta0}}\label{app:proofs-proposition}
For $n = 0$ the recursion follows trivially. 
For $n > 0$, at each iteration, we can either have 
\begin{align*}
    \mathcal{Q}_n &= \max_{\theta} \sum_{t = 1}^n h(x_t, \theta) - \max_\theta Q_n(\theta) = \\
    &= \max_{\theta} \sum_{t = 1}^n h(x_t, \theta) - \max_{\theta} \sum_{t = 1}^{n} h(x_t, \theta),
\end{align*}
if there is no change up to time $n$, or for a change at time $\tau \in 1, \dots, n-1$:
\begin{align*}
    \mathcal{Q}_n &= \max_{\theta} \sum_{t = 1}^n h(x_t, \theta) - \max_\theta Q_n(\theta) = \\
    &= \max_{\theta} \sum_{t = 1}^n h(x_t, \theta) - \max_\theta\left[ Q_{n-1}(\theta) + h(x_n, \theta) \right] =\\
    &= \max_{\theta} \sum_{t = 1}^n h(x_t, \theta) -\\
    & \max_\theta\left[ \max_{\theta_0}\sum_{t = 1}^{\tau} h(x_t, \theta_0) + 
    \sum_{t = \tau+1}^{n-1} h(x_t, \theta) + h(x_n, \theta) \right].
\end{align*}
\hfill $\Box$
\vspace{5pt}

\subsection*{Proof of Theorem \ref{theo:funcbound}}\label{app:proofs-theo3}
The proof follows almost exactly the one derived for Theorem 3 in \cite{romano2021fast}. While they assume a piecewise constant process plus \textit{i.i.d.} noise from a continuous distribution, we denote how this assumption can be relaxed to include any exchangeable real-valued random process with an additive cost function, such as our Bernoulli process. 
Their proof relies on the following lemmas. 

    \begin{lemma}\label{lem:inclusion2}
    For $i \leq j \leq k$
    \begin{equation}
    \mathcal{I}_{i:k} \quad \subseteq \quad \mathcal{I}_{i:j}  \cup     \mathcal{I}_{j+1:k}
    \end{equation}
    \end{lemma}
    
    \begin{lemma}\label{lem:convexhull}
    The set of $\tau$ in $\mathcal{I}_{i:j}$ are the extreme points of the largest convex minorant of the sequence $S_{i:j} = \{\sum_{t = i}^i x_t, \dots,  \sum_{t = i}^j x_t\}$. 
    \end{lemma}

We notice that if we write the cost of a segment with a change at $\tau \in \mathbb{N}$ as
\begin{equation*}
    \begin{split}
    q_{i:j, \tau}(\theta_0, \theta_1) = \sum_{t=i}^{\tau} \left[x_t \log\theta_0 + (1-x_t)\log(1-\theta_0) \right] +\\ \sum_{t=\tau+1}^j \left[x_t \log\theta_1 + (1-x_t)\log(1-\theta_1) \right],        
    \end{split}
\end{equation*}
then for any $\tau, \tau'$: 
\begin{equation*}
    \begin{split}
    &q_{i:j, \tau}(\theta_0, \theta_1) - q_{i:j, \tau'}(\theta_0, \theta_1) =\\ 
    &=\sum_{t = \tau}^{\tau'} \left[ x_t \log\left(\frac{\theta_1}{\theta_0}\right) + (1-x_t) \log\left(\frac{1-\theta_1}{1-\theta_0}\right) \right]        
    \end{split}
\end{equation*}
this does not depend on $i$ and $j$. This allows us to extend Lemmas \ref{lem:inclusion2} and \ref{lem:convexhull} to the Bernoulli case.
    
And as both \cite{andersen1955fluctuations} and \cite{abramson2012some} assume an exchangeable real-valued random variable, we can adapt Lemma 7 of \cite{romano2021fast} to the Bernoulli case:
    
\begin{lemma}\label{lemma:Andersen}
    Assuming the $x_t$ follows an exchangeable real-valued process $\forall t \in i:j$, then 
    $E(\#\mathcal{I}_{i:j})=\sum_1^{j-i-1} 1/(t+1)$.
\end{lemma}
    
Having shown that the three lemmas apply to the Bernoulli case we can simply follow the proof from \cite{romano2021fast}.
\hfill $\Box$

\subsection{Detailed Description of Simulation Scenarios}\label{sec:sim_descr}

In this section we describe the simulation scenarios, detailing the sampling for respectively the pre-change and post-change distributions. 
The Cauchy change-in-scale was obtained by sampling $y_{1:\tau}$ from a Cauchy with location 0 and scale 1, \textit{i.e.}, $y_{1:\tau} \sim \textit{Cauchy}(0, 1)$ and for post change $y_{\tau+1:n} \sim Cauchy(0, 5)$. Following similar notation, the Gaussian change-in-mean was obtained by $y_{1:\tau} \sim N(0, 1)$ and $y_{\tau+1:n} \sim N(1, 1)$. 
The multimodal scenario was obtained by sampling $y_t \sim N(0, 1)$ with probability $\alpha$ and $y_t \sim N(10, 1)$ with probability $(1-\alpha)$, for $\alpha = 2/3$, $t = 1, \dots, \tau$ and for $\alpha = 1/3$, $t = \tau + 1, \dots, n$ after the change.

In the OU scenario, the data is generated by simulating a discrete Ornstein-Uhlenbeck process (also known as mean-reverting random walk) plus additional Gaussian noise. This is $y_t = \nu_t + \epsilon_t$, where $\epsilon_t \sim N(0, \sigma_\epsilon^2)$ is a white noise process and $\nu_t$ is an autoregressive process defined by the recursion $\nu_t = \nu_{t-1} - \theta f_{t-1} - \theta \nu_{t-1} + \sigma_\nu w_{t-1}$, where $\theta = 0.1,\ w_t \sim N(0, \sigma_\eta^2)$. The sequence $f_t$ is added to encode a shift in the mean of the Ornstein-Uhlenbeck process: for $t = 1, \dots, \tau$, $f_t = 0$, and for $t = \tau + 1, \dots, n$, $f_t = -10$.

The sinusoidal scenario is generated by sampling $y_t \sim N(\mu_t, 1)$ for $t = 1, \dots, n$. The mean $\mu_t$ is given by the function $\mu_t = \sin(\pi f t)$, where $f = 0.2$ is the frequency of the sinusoidal function. From the changepoint $\tau$, the mean starts to decrease exponentially with rate $\lambda = 0.005$, \textit{i.e.} for $t = \tau + 1, \dots, n$, $\mu_t = A \sin(\pi f t) \exp(-\lambda t)$. 

Finally, the change in tails scenario is generated by sampling $y_t \sim t_\nu $ for $t = 1, \dots, n$, where $t_\nu$  is a Student's t-distribution with $\nu$ degrees of freedom. After the changepoint, for $t = \tau + 1, \dots, n$, we randomly selected a $5^{th}$ of the observations and increase by a random amount drawn from a Poisson distribution with mean 10.

\subsection{Additional Simulation Results}\label{sec:add_empirical}

In Section \ref{fig:choice_of_quantiles} we have commented on the choice of $M$, the number of quantiles. As for the quantile points, when those cannot be provided directly by the practitioners, those needs to be estimated through a probation period. We therefore performed a simulation study to see the effect of the probation period (amount of training observations) for the quantile estimation on the performances of NP-FOCuS. As in the main study, we present detection delay and false positive rate in Tables \ref{tab:prob_det_delay} and \ref{tab:prob_fpr}. We found that, in general, even as little as 100 observations are enough for estimating the quantiles. In many cases the length of the probation period has marginal impact on the performances. The major differences are seen on the sinusoidal scenario, where 25 and 50 observations are not enough to capture the full domain of the distribution, resulting in significantly more false positives or slower detections.

\begin{table*}[ht]
\centering
\begin{subtable}{\textwidth}
\centering
\begin{tabular}{llllll}
  \hline
  scenario & 25 & 50 & 100 & 150 & 200 \\ 
  \hline
 OUmean & 102.97 & 94.29 & 87.99 & \textbf{78.46} & 80.71 \\ 
   cauchy & 46.19 & 46.45 & \textbf{33.98} & 34.47 & 43.82 \\ 
   gauss & 29.57 & 28.3 & \textbf{22.26} & 27.4 & 26.58 \\ 
   multim & 71.6 & 61.24 & \textbf{44.86} & 55.9 & 60.88 \\ 
   sinusoidal & 154.68 & 252.3 & 165.8 & 140.92 & \textbf{130.61} \\ 
   tails & \textbf{36.55} & 55.85 & 46.97 & 49.33 & 44.4 \\ 
   \hline
\end{tabular}
    \caption{Detection delay }
    \label{tab:prob_det_delay}
\end{subtable}

\vspace*{1 cm} 

\begin{subtable}{\textwidth}
\centering
\begin{tabular}{llllll}
  \hline
  scenario & 25 & 50 & 100 & 150 & 200 \\ 
  \hline
 OUmean & 0 & 0 & 0 & 0.01 & 0.01 \\ 
   cauchy & 0 & 0 & 0.01 & 0.01 & 0 \\ 
   gauss & 0 & 0 & 0.01 & 0 & 0 \\ 
   multim & 0.01 & 0.01 & 0.03 & 0.02 & 0.01 \\ 
   sinusoidal & 0.08 & 0.03 & 0.03 & 0.04 & 0.05 \\ 
   tails & 0.02 & 0 & 0 & 0 & 0 \\ 
   \hline
\end{tabular}
    \caption{False positive rate}
    \label{tab:prob_fpr}
\end{subtable}
\caption{Average detection delay and false positive rate for all scenarios of our simulation study over a range of different probation periods.}
\end{table*}

\begin{figure}[tb]
    \centering
    \includegraphics[width=\linewidth]{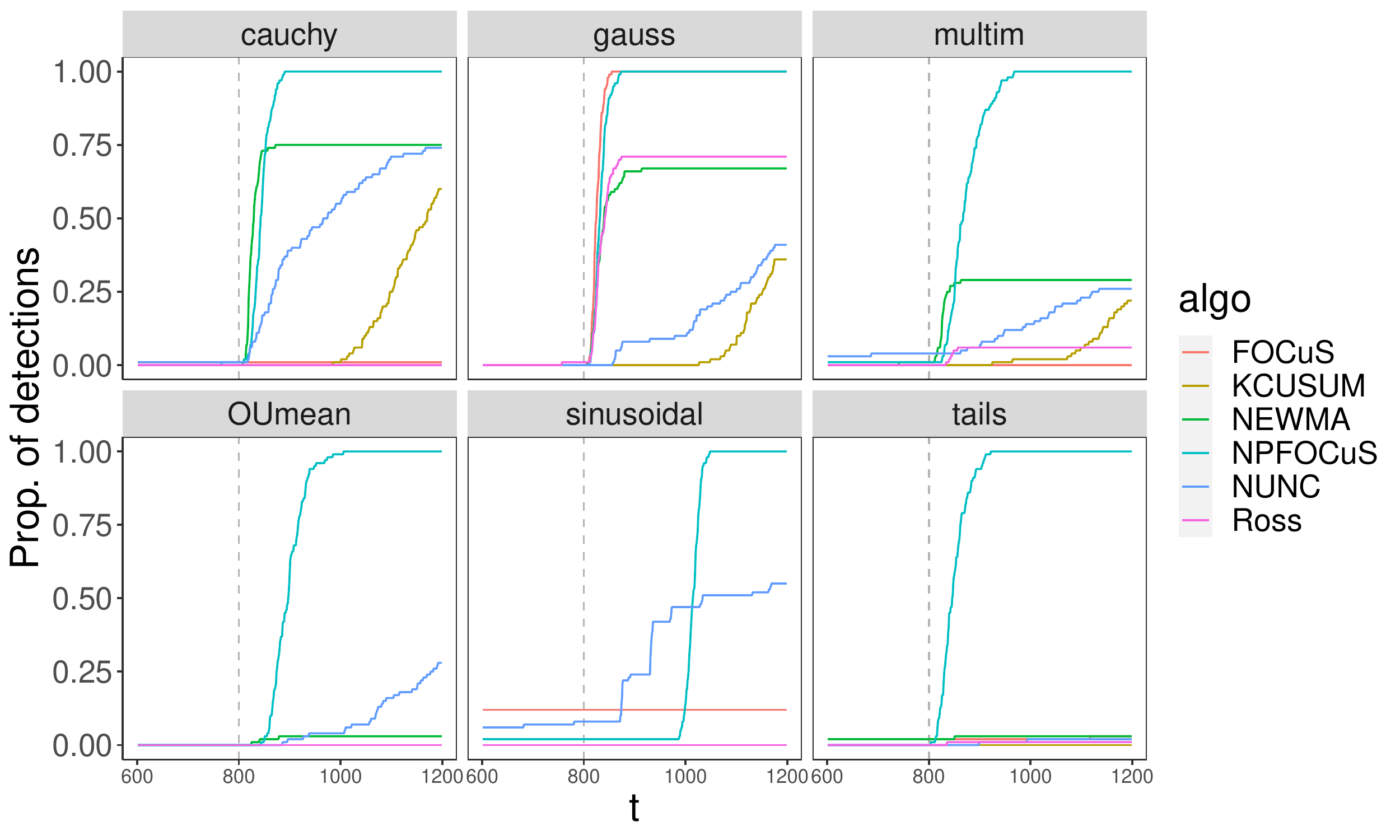}
    \caption{Proportions of changes detected within $t$ observations following the change in six different change scenarios. The change is denoted by the vertical dotted line at $t = 800$.}
    \label{fig:prop-of-detections-800}
\end{figure}

To see how the methods deal with an earlier change, we replicate the study of Section \ref{sec:simulations}, but we shift the change at time $\tau = 800$. Results are summarised in Figure \ref{fig:prop-of-detections-800}.
We denote how there is a significant drop in performances in NEWMA across all simulations scenarios. This is particularly evident in the Cauchy and multimodal scenario. The reduction in detection power is attributable to a smaller number of random features, in order to stabilise the statistics faster and achieve shorter burn-in periods. This result is in line with what observed in Section VI-B of \cite{kerivenNEWMA}.

\vspace{5pt}

\end{document}